# Stimulated emission of plasmon-LO mode in narrow gap HgTe/CdHgTe quantum wells

V Ya Aleshkin[1,2], A A Dubinov[1,2], V I Gavrilenko[1,2] and F Teppe[3]

[1] Department of Semiconductor Physics, Institute for Physics of Microstructures RAS, Nizhny Novgorod, 603950, Russia
[2] Lobachevsky State University of Nizhny Novgorod, Nizhny Novgorod, 603950, Russia
[3] Laboratoire Charles Coulomb, Université de Montpellier, Centre National de la Recherche Scientique, 34095 Montpellier, France

E-mail: aleshkin@ipmras.ru

**Abstract**

The possibility of amplification of the coupled plasmon – optical phonon (LO) mode in HgTe quantum wells (QWs) is theoretically considered assuming an inverted band population. We calculate the dispersion of the plasmon–LO modes taking into account the spatial dispersion of the electronic polarizability. It is shown that stimulated emission of the plasmon – LO mode is possible in the frequency range corresponding to the *Reststrahlen* band of GaAs both in a 6-nm-wide HgTe/CdTe QW and in a 5-nm-wide HgTe/Cd$_{0.7}$Hg$_{0.3}$Te QW grown on the (013) plane. Due to the anisotropy of the dispersion law for the plasmon – LO mode, the [03-1] direction appears to be optimal for generation.

## 1. Introduction

Terahertz (THz) frequency range still lacks compact and effective radiation sources demanded for numerous applications. Unipolar quantum cascade lasers (QCLs) based on A$_3$B$_5$ semiconductors nowadays demonstrate remarkable performance in between 1 and 5 THz (λ = 60 – 300 μm) and above 15 THz (λ < 20 μm) [1]. However their characteristics drop at λ > 20 μm due to strong lattice (two-phonon) absorption and in 20-28 μm range QCL operation has been demonstrated at several wavelengths only (see [2] and references therein). The spectral range up to λ = 50 μm is covered by PbSnSe interband diode lasers [3]. In this material, low energies of optical phonons (compared with that in A$_3$B$_5$ semiconductors) result in two-phonon absorption shifting to longer wavelengths while the symmetry of electron and hole energy-momentum laws suppresses the nonradiative Auger recombination. However, the characteristics of lead salt lasers are largely limited by the growth technology. The abovementioned symmetry of electron and hole energy-momentum laws can also be achieved in HgTe/CdHgTe quantum wells (QWs) and should facilitate the mitigation of Auger processes [4]. Since transverse optical phonons in HgCdTe have low energies compared to A$_3$B$_5$ semiconductors, lasing in HgTe/HgCdTe QW heterostructures could be possible up to wavelength about 50 μm [5]. To the moment the stimulated emission in HgTe/CdHgTe QW heterostructures with dielectric waveguides has been demonstrated at wavelengths 19.5 μm [6] and 24 μm [7]. However, for longer wavelengths it is necessary to compensate the increase in two-phonon absorption and to grow thick (over 20 μm) HgCdTe waveguide structure, which is challenging for molecular beam epitaxy technology.

The generation of two-dimensional plasmons in the terahertz region of the electromagnetic spectrum in narrow-gap or gapless semiconductors (with subsequent re-emission of photons) could be a promising alternative to the stimulated emission of photons. The first advantage is a large plasmon gain arising from high level localization of electromagnetic fields of two-dimensional plasmon near quantum well. The second major advantage is that one needs no special waveguide structure, which introduces a noticeable part of losses in terahertz cascade lasers [1, 8]. To date, comprehensive theoretical studies are available on the generation of plasmons in graphene [9-12]. However, the recombination times in graphene are extremely short – on the order of several picoseconds. Short recombination times complicate achieving the population inversion of the bands, necessary for the coherent generation of plasmons.

Recently, it was proposed to generate two-dimensional plasmons in narrow-gap HgTe/CdHgTe quantum wells [13]. In these structures, the lifetimes of nonequilibrium carriers are several orders of magnitude longer than in graphene. In addition, the technology of their growth, in contrast to graphene, is quite well developed [14]. Therefore, HgTe/CdHgTe QWs seem to be the attractive candidates for the sources of coherent two-dimensional plasmons, especially in the GaAs *Reststrahlen* band (photon energy range 25 to 50 meV), in which QCLs face formidable problems [1]. The main features and peculiarities of the generation of two-dimensional plasmons in HgTe/CdHgTe QWs were outlined in [13]. However, Ref [13] implemented a simplified model leaving aside some important factors for the problem of plasmon generation. The first factor is the spatial dispersion of the polarizability in the electron and hole gases, which is essential for finding the correct dependence of the plasmon frequency on its wave vector $q$. As shown below, taking this dispersion into account significantly changes the dependence of the frequency of a two-dimensional plasmon on its wave vector in the region where plasmons are amplified due to interband electron transitions (see also [15]). In particular, one yields $\omega \sim \sqrt{q}$ when neglecting the spatial dispersion in this region [13], while when the dispersion is taken into account the frequency dependence on the wave vector becomes linear $\omega \sim q$. The second factor is the anisotropy of the electron and hole dispersion laws that leads to the anisotropy of the dispersion law and the amplification coefficients for two-dimensional plasmons. Finally, two other important factors are the dispersion of the dielectric constant due to optical lattice vibrations and losses due to phonon absorption. These factors are important for the generation of plasmons in the frequency range corresponding to the GaAs *Reststrahlen* band. For example, the dispersion of the dielectric constant due to optical lattice vibrations leads to the formation of the coupled plasmon-LO modes.

This work is devoted to the theoretical consideration of the coherent generation of plasmon-LO modes in narrow-gap HgTe/CdTe QW heterostructures taking into account the factors mentioned above. Using 6-nm-wide and 5-nm-wide QWs for illustration, we calculate the dispersion law of the plasmon-LO modes, taking into account the spatial dispersion of the polarizability of electrons and holes and the frequency dispersion of the refractive index caused by optical vibrations of the lattice. It is shown that stimulated emission of the plasmon – LO mode is possible in the frequency range corresponding to the *Reststrahlen* band of GaAs both in a 6-nm-wide HgTe/CdTe QW and in a 5-nm-wide HgTe/Cd$_{0.7}$Hg$_{0.3}$Te QW grown on the (013) plane. Due to the anisotropy of the dispersion law for the plasmon – LO mode, the [03-1] direction appears to be optimal for generation.

## 2. Calculation model

To find the plasmon dispersion law, it is necessary to calculate the polarizability of two-dimensional electron and hole gases. Generalizing the Linhard formula [16], one can obtain the following expression for the polarizability of a two-dimensional electron-hole system (see Appendix 1):

$$\chi(q,\omega) = \frac{e^2}{q^2(2\pi)^2} \times$$
$$\sum_{s,s'} \int d^2k \frac{[f_s(\mathbf{k}) - f_{s'}(\mathbf{k}+\mathbf{q})] | \psi^+_{\mathbf{k}+\mathbf{q},s'} \psi_{\mathbf{k},s} |^2}{\varepsilon_{s'}(\mathbf{k}+\mathbf{q}) - \varepsilon_s(\mathbf{k}) - \hbar\omega - i\hbar\nu_{s'\mathbf{k}+\mathbf{q};s,\mathbf{k}}} \quad (1)$$

where the lower index denotes the number of the subband including the spin index, $\varepsilon_s(\mathbf{k})$ is the energy of an electron with a wave vector $\mathbf{k}$ in the $s$-th subband, $f_s(\mathbf{k})$ is the electron distribution function in the corresponding subband, -$e$ is the electron charge, $\nu_{s,s'}$ are the phase relaxation frequencies for the density matrix component $\rho_{s's}(\mathbf{k}+\mathbf{q},\mathbf{k})$, $\hbar$ is the reduced Plank's constant. The wave function of an electron in the $s$-th subband has the form $\psi_{\mathbf{k},s}(\mathbf{r}) = \psi_{\mathbf{k},s}(z)\exp(i\mathbf{kr})/\sqrt{S}$, where $z$ is the coordinate along the normal line to the plane of the quantum well, $S$ is the area of the quantum well.

$$| \psi^+_{\mathbf{k}+\mathbf{q},s'} \psi_{\mathbf{k},s} | = | \int dz \psi^+_{\mathbf{k}+\mathbf{q},s'}(z) \psi_{\mathbf{k},s}(z) | \quad (2)$$

Note that in Eq. (1), the intersubband contributions to the polarizability that were neglected in [13, 15] are taken into account. In our calculation the electron distribution functions were assumed to be as follows:

$$f_s(\mathbf{k}) = \left(1 + \exp\left(\frac{\varepsilon_s(\mathbf{k}) - F_s}{k_B T_e}\right)\right)^{-1} \quad (3)$$

where $F_s$ is the chemical potential in the $s$-th subband ($F_s = F_c$ in the conduction band and $F_s = F_v$ in the valence band), $k_B$ is the Boltzmann constant, $T_e$ is the effective temperature of electrons and holes, which can differ from the lattice temperature. The values $F_{c,v}$ were found from the condition of equality for the concentration of electrons and holes, calculated using the expression (3) for the given concentrations.

The dependence of the plasmon frequency $\omega$ on its wave vector $q$ can be found from the following relation [17]:

$$1 + \frac{2\pi}{\kappa(\omega)} \chi(q,\omega) \sqrt{q^2 - \left(\frac{\omega}{c}\right)^2 \kappa(\omega)} = 0, \quad (4)$$

which can be obtained from Maxwell's equations. In (4), $c$ is the velocity of light in vacuum, $\kappa(\omega)$ is the dielectric permittivity of the HgCdTe barriers. It should be noted that (4) is valid when the width of the quantum well is much less than $1/q$. Note that in (1) and (4), the plasmon wave vector $q$ is assumed to be real, and its frequency $\omega$ is assumed to be complex.

To describe the dielectric permittivity of CdTe, we will use the well-known expression [18]:

$$\kappa(\omega) = \kappa_\infty \left(1 + \frac{\omega_L^2 - \omega_T^2}{\omega_T^2 - \omega^2 - i\Gamma\omega}\right). \quad (5)$$

For CdTe, $\kappa_\infty = 7.1$, $\omega_L = 169$ cm$^{-1}$, $\omega_T = 141$ cm$^{-1}$, $\Gamma = 6.6$ cm$^{-1}$ [19]. Expression (5) does not describes two- and three-phonon absorption. Absorption is determined by the imaginary part of the dielectric permittivity. It is worth comparing the dependence of the imaginary part of the dielectric permittivity of formula (5) and the experimentally measured value. The experimentally measured dependences of the real and imaginary parts of the refractive index on the wavelength can be found in the book [19]. Using this data, we calculated the imaginary part of the dielectric constant. The dependences of the imaginary and real parts of the dielectric permittivities on the photon energy of expression (5) and obtained from the processing of experimental results are given in Fig.1. One can be see that the imaginary part $\kappa(\omega)$ of expression (5) is larger than the experimentally obtained one in the entire considered spectral range, with the exception of a narrow region around 36 meV. Therefore, using (5), we will not overestimate the plasmon gain. The results for the real part of the permittivity are in good agreement for expression (5) and the experimentally measured values. In the calculations, it is more convenient to use the analytical expression (5), rather than the experimentally measured values of the dielectric permittivity.

As for as dielectric permittivity of the solid solution Cd$_x$Hg$_{1-x}$Te is concerned, in a number of studies [20-22] it was found that it can be represented in the form [21]:

$$\kappa(\omega) = \kappa_\infty + \sum_{j=1}^{N} \frac{S_j \omega_{jT}^2}{\omega_{jT}^2 - \omega^2 - i\Gamma_j \omega} \quad (6)$$

where the number $N$ is selected from considerations of fitting to experimentally obtained reflection spectra. For example, for a Cd$_{0.7}$Hg$_{0.3}$Te solid solution, the sum on the right-hand side of (6) with eight terms [22] is used to describe the dielectric permittivity. We used the values $\kappa_\infty$, $S_j$, $\omega_{jT}^2$, $\Gamma_j$ from this work to describe plasmons in a 5 nm wide HgTe/Cd$_{0.7}$Hg$_{0.3}$Te QW.

The electron and hole energy spectra in the QWs were calculated using the 4-band Kane model (Hamiltonian 8x8) with allowance for deformation effects. For simplicity, we did not take into account the effects due to the absence of an inversion center in the crystal lattice and a decrease in symmetry at heterointerfaces, which remove the spin degeneracy. Details of the calculations can be found in [23]. We consider the HgTe/Cd$_x$Hg$_{1-x}$Te structures grown on the (013) plane, since the main achievements in the field of stimulated emission generation in recent years have been obtained just on such structures [6, 24, 25].

Since this work is devoted to calculating the gain of the plasmon-LO mode, we will consider a nonequilibrium situation when the hole concentration is equal to the electron concentration, i.e. the concentration of photoexcited carriers is much higher than that of equilibrium ones.

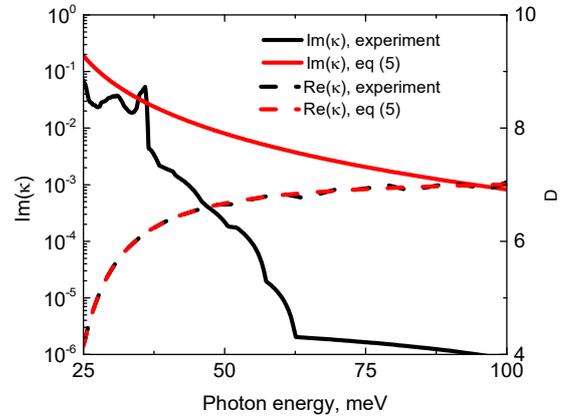

Fig.1. Dependences of the real and imaginary parts of the dielectric permittivity of CdTe on the photon energy. Red lines correspond to expression (5), black – to experimental results [19].

## 3. Results and discussion

Calculated energy spectra of electrons and holes in a 6 nm wide HgTe/CdTe QW and in a 5 nm wide HgTe/Cd$_{0.7}$Hg$_{0.3}$Te QW are shown in Fig.2. In both cases the gaps between the lowest conduction E1 and top valence (HH1) subbands (30 meV in Fig.2a and 35 meV in Fig.2b) are much less than the distance between E1 and the second conduction subband E2 at k = 0 (347 meV and 344 meV respectively). Therefore, for calculating the spectra of the plasmon-LO modes, the high-lying conduction subband can be neglected. However, when calculating the gain of the plasmon-LO modes in the structures under consideration at high carrier concentrations, it is necessary to take into account the second hole subband HH2. At high excitation level hole populate in HH1 subband the states close to the minimal distance between HH1 and HH2 subbands (k ~ 0.2 nm$^{-1}$). Therefore the electron transitions between HH2 and HH1 subbands with plasmon absorption (shown by arrows in Fig.2) become possible that should be taken into account at the calculations amplification (absorption) of the plasmon-LO modes.

*3.1 Dispersion of plasmon-LO modes*

As noted in the Introduction, due to the interaction of plasmons with optical lattice vibrations in binary semiconductors, two coupled plasmon-LO phonon modes appear. For three-dimensional plasmons, a similar phenomenon is well known (see, for example, problem 6.10 from book [18]). For simplicity, consider first a HgTe well sandwiched between CdTe barriers. In this case, the dispersion law of the low-frequency mode at small q is close to the square root law characteristic of a two-dimensional plasmon [13], when the spatial dispersion of the polarizability is insignificant. The energy of this mode does not exceed the energy of the transverse optical (TO) phonon in CdTe (18.4 meV). If there were no Landau damping (absorption of the plasmon-LO mode due to intraband electron transitions), then with increasing $q$ the energy of this mode would tend to the energy of the TO phonon. The energy of the high-frequency mode tends to the LO phonon energy in CdTe (21 meV) as $q$ tends to zero. With an increase in the wave vector, the dispersion law of the high-frequency mode is close to linear (see Fig. 3). In the *Reststrahlen* band (the region between the energies of TO and LO phonons), plasmon modes are absent.

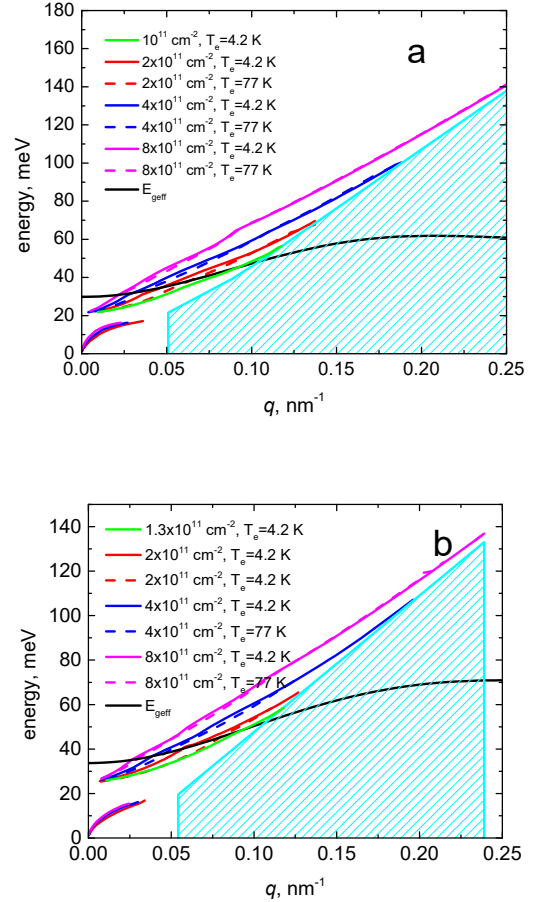

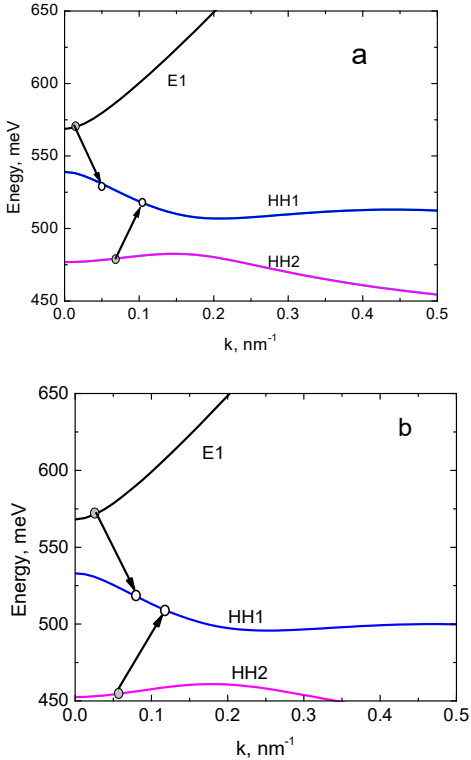

Fig. 3. Dependences of the energy of the high-frequency plasmon-LO mode on the wave vector at different concentrations and different electron temperatures of charge carriers in a 6 nm wide HgTe/CdTe QW (a) and 5 nm wide HgTe/Cd$_{0.7}$Hg$_{0.3}$Te QW (b); lattice temperature T = 4.2K. The wave vector $q$ is directed along the [100] direction. In the shaded region (abrupted at $q$ ~0.05 nm$^{-1}$) there is intraband absorption of the plasmon-LO modes (Landau absorption). The black line denotes the dependence of the effective band gap on the wave vector for interband electron transitions with the emission of the plasmon-LO mode quanta.

Since the energy of the low-frequency mode is less than the band gap of the QWs under consideration, amplification of this mode is impossible when the bands are inversely populated by nonequilibrium carriers. For this reason, we will not consider it, but restrict ourselves to considering the high-frequency plasmon - LO phonon mode. Sometimes we will use the term "plasmon" to denote the quantum of this mode.

Fig. 2 Energy spectra in a 6 nm wide HgTe/CdTe QW (a) and in a 5 nm wide HgTe/Cd$_{0.7}$Hg$_{0.3}$Te QW (b); T = 4.2 K. The transitions with plasmon emission and absorption are show by down and arrows up, respectively.

In the case of barriers of Cd$_x$Hg$_{1-x}$Te solid solution, the energy of the high-frequency mode tends to the energy of the highest-frequency phonon longitudinal mode in Cd$_x$Hg$_{1-x}$Te as $q$ tends to zero. This statement is easy to prove if one does not take into account the imaginary part of the dielectric constant of the lattice and the spatial dispersion of the polarizability, which is insignificant at small values of $q$.

The calculated spectra of the high- and low-frequency modes at various concentrations of nonequiilibrium electrons and holes at electronic temperatures $T_e = 4.2$ K and $T_e = 77$ K in 6 nm and 5 nm wide QWs are shown in Fig.3. The phase relaxation frequencies necessary for the calculation were estimated from the typical values of the electron and hole mobilities [26]. In our calculations, we assumed $\hbar \nu_{ss'} = 1$ meV if both indices $s, s'$ correspond to the conduction band, $\hbar \nu_{ss'} = 2$ meV, if both indices $s, s'$ correspond to the valence band. In the case when the indices $s, s'$ correspond to different bands, it was assumed $\hbar \nu_{ss'} = 1.5$ meV. Note that a decrease in these values (or an increase by a factor of 2–3) has practically no effect on the dispersion law for the plasmon-LO modes.

Figure 3 shows that with an increase in the wave vector, the dispersion law of the high-frequency mode is close to linear. As the carrier concentration increases, the slope of the high-frequency mode dependence increases (i.e., the group and phase velocities of the plasmon-LO mode increase). The characteristic values of the group velocities of the high-frequency mode are $\sim 5 \cdot 10^7$ cm/s for $n$ (electron concentration) = $p$ (hole concentration) = $2 \cdot 10^{11}$ cm$^{-2}$ and $\sim 7 \cdot 10^7$ cm/s for $n = p = 8 \cdot 10^{11}$ cm$^{-2}$, respectively.

In Fig. 3, in the shaded region, absorption of the high-frequency plasmon-LO mode is possible due to intraband transitions (Landau absorption). In this region, the plasmon-LO mode is strongly damped, and therefore we do not consider it there. The upper boundary of this region at $T_e = 0$ is determined by the absorption of the plasmon-LO mode by electrons of the conduction band with the Fermi energy [13]; it was found from the equation $\hbar \omega_{max}(q) = \varepsilon_c(k_F + q) - \varepsilon_c(k_F)$, where $\varepsilon_c(k)$ is the dependence of the electron energy in the conduction band on its wave vector k. With increasing temperature $T_e$, the boundary of the Landau damping region for a fixed wave vector shifts to higher frequencies.

Note one important feature when considering interband electronic transitions at absorption or emission of the plasmon. Due to the large effective refractive index of the plasmon waveguide, the wave vector of the plasmon can be comparable to the wave vectors of an electron in the initial and final states at an interband transition (a diagrams of such transitions are shown in Fig.2 by arrows). Therefore, when considering such processes, one cannot neglect the wave vector of the plasmon, as is usually done for interband electron transitions with absorption (emission) of a photon in semiconductors [18]. This circumstance leads to the fact that the minimum energy of the plasmon, which an electron can emit during the transition from the conduction band to the valence band, depends on the wave vector $q$ [13]. This dependence is shown in Fig. 3 with black lines and denoted by $E_{geff}(q)$. In the vicinity of the crossing of the dispersion law of the plasmon-LO mode with $E_{geff}(q)$, there is a small kink in the dependence $\hbar\omega(q)$ is due to the "inclusion" of interband terms in the polarizability due to electronic transitions between the conduction and valence bands. A similar feature in the dependence $\hbar\omega(q)$ is clearly seen in Fig. 3a for a carrier concentration of $8 \cdot 10^{11}$ cm$^{-2}$ at $q \sim 0.9$ nm$^{-1}$. However, this feature is due to electron transitions between the valence subbands (see Fig. 2). Note that an increase in the effective carrier temperature $T_e$ has a small effect on the dispersion law of the plasmon-LO mode.

At a carrier concentration below $10^{11}$ cm$^{-2}$ at $T_e = 4.2$ K in a 6 nm wide HgTe/CdTe QW, the high-frequency mode dispersion in the unshadowed region in Fig.3a is located below the black line corresponding to $E_{geff}(q)$. In this case, amplification of the plasmon-LO mode is prohibited by the energy and momentum conservation laws for interband electronic transitions with plasmon emission. Therefore, there is a certain critical carrier concentration below which the plasmon emission does not take place [13]. In particular, for a 6 nm wide HgTe/CdTe QW, this concentration is approximately equal to $10^{11}$ cm$^{-2}$ at $T_e = 4.2$ K when the plasmon-LO mode propagates along the [100] direction.

Let us note two more important circumstances. In [13], the spatial dispersion of the polarizability of carriers in the QW and the interaction of plasmons with LO phonons were not taken into account. Therefore, in this work, the plasmon dispersion law at all frequencies was taken as follows:

$$\omega(q) = \sqrt{\frac{2\pi |q|}{\kappa_\infty}\left(\frac{n}{m_c} + \frac{p}{m_v}\right)} \quad (7)$$

Such dispersion is quite different from that of high-frequency plasmon-LO modes shown in Fig.3, for which plasmon emission is possible. In work [15], the frequency dispersion of the polarizability was allowed for, but the interaction with LO phonons was not taken into account. However, at frequencies significantly exceeding the LO phonon frequency, this interaction weakly affects the plasmon dispersion law. More precisely, the frequency dispersion of the polarizabilities of electrons and holes was taken into account in [15] only for intraband transitions (i.e., $s = s'$ was set in (1)). The last approximation changes the dispersion law of the plasmon-LO modes within 10%.

### 3.2 Plasmon amplification

To calculate the gain (absorption) value, it is necessary to find the imaginary part of the wave vector at the real value of the frequency. In our approach, we find the complex frequency at a real wave vector. Let us show how to find the gain from this dependence. In general, on the complex plane of wave vectors, frequency is a complex function. We have obtained the complex frequency for a real wave vector $q_0$. Frequency is an analytical function, so its dependence on the wave vector in the vicinity $q_0$ can be represented in the following form, using the Taylor series expansion:

$$\omega(q) \approx \omega(q_0) + v_g(q_0)(q - q_0), \quad v_g(q) = \frac{d\omega}{dq}, \quad (8)$$

where $v_g(q)$ is the complex group velocity. In the vicinity $q_0$, we find a point $q$ at which the following condition will be satisfied: the frequency must be equal to the real part $\omega(q_0)$, i.e. $\omega(q) = \mathrm{Re}(\omega(q_0))$. Note that the imaginary part of the frequency at $q_0$ is small in comparison with the real part of the frequency (the difference is more than an order of magnitude). Therefore, in expansion (8), we can restrict ourselves to the first two terms of the Taylor series. From the condition imposed on $q$, from (8) it follows that $v_g(q_0)(q - q_0) = -i\,\mathrm{Im}(\omega(q_0))$. From this equality we find $\mathrm{Im}(q) = -\mathrm{Im}(\omega(q_0))\mathrm{Re}(v_g^{-1})$. Since the absorption coefficient $\alpha_{abs}$ is equal to twice the imaginary part of the wave vector, the gain as a function of frequency can be written as:

$$\begin{aligned}\alpha_{amp}(\mathrm{Re}(\omega(q_0))) &= -\alpha_{abs}(\mathrm{Re}(\omega(q_0))) \\ &= 2\,\mathrm{Im}(\omega(q_0))\,\mathrm{Re}\left(v_g^{-1}(q_0)\right)\end{aligned} \quad (9)$$

Calculated spectra of the gain of the plasmon-LO mode at different concentrations of charge carriers and electron temperatures are given in Fig.4. As easy to see with an increase in the carrier concentration, the gain spectrum becomes broader due to the shift of its high-energy boundary. This mainly results from the increase with the carrier concentration of the energy region in which an inverted population exists. The low-energy boundary of the gain is determined by the crossing of the dispersion curve of the plasmon-LO mode and effective band gap $E_{g\mathrm{eff}}(q)$ curve (see Fig.3). As can be seen from Fig. 3 and Fig. 4, for carrier concentrations $\geq 2\cdot 10^{11}$ cm$^{-2}$, the low-energy boundary of the amplification region is close to the band gap (slightly more than $E_g = 30\text{-}35$ meV), and for a concentration of $10^{11}$ cm$^{-2}$ it is about 40 meV (Fig.4a).

The gain spectra for carrier concentrations of $4\cdot 10^{11}$ cm$^{-2}$ and $8\cdot 10^{11}$ cm$^{-2}$ in a 6 nm wide HgTe/CdTe QW clearly show features (dips) at energies of about 40 and 60 meV (Fig.4a). These features are due to hole transitions between the valence subbands. These features disappear if the contribution from the excited valence subband is excluded from the expression for the polarizability (1). The higher the concentration of holes in the valence band, the more pronounced these features are. For example, for a carrier concentration $8\cdot 10^{11}$ cm$^{-2}$ at $T_e = 77$ K, one can see Fig. 4 two amplification spectral regions, separated by an absorption region resulted from the transitions of electron transitions between the hole subbands HH2 and HH1.

Note that the "critical" concentration at which the inverted population of the bands and plasmon amplification are realized increases with $T_e$ rise. Thus at $T_e = 4.2$ K in a 6 nm wide HgTe/CdTe QW it is close to $10^{11}$ cm$^{-2}$ and at $T_e = 77$ K it is approximately equal to $3\cdot 10^{11}$ cm$^{-2}$. Figure 4 shows that the magnitudes of the gain at $n = p = 2\cdot 10^{11}$ cm$^{-2}$ can exceed $10^5$ cm$^{-1}$, which is due to the low group velocity of the plasmon-LO mode and its strong localization near the QW. Such values of the gains open up the possibility of coherent generation of the plasmon-LO mode in short (~ 1 μm) samples.

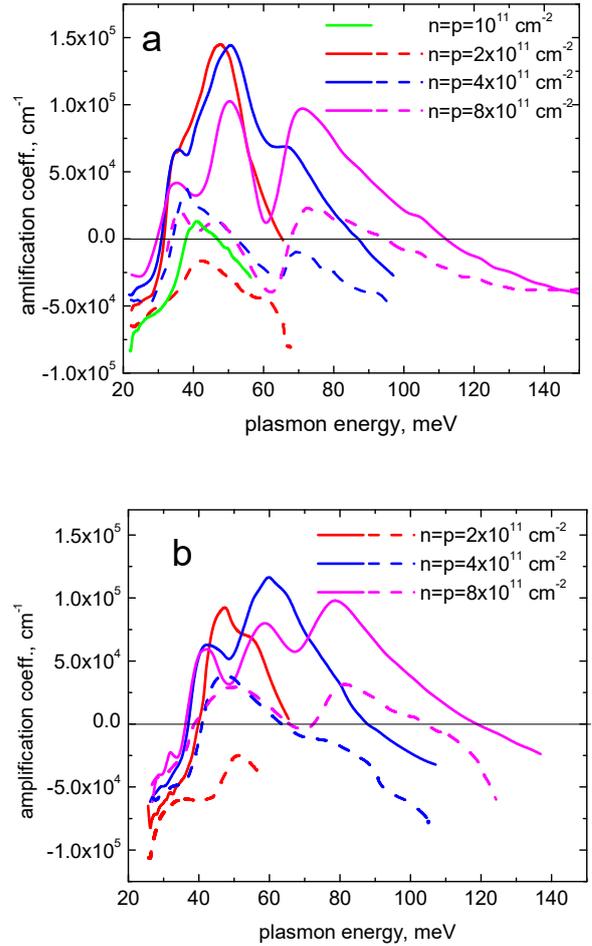

Fig. 4. Calculated gain spectra of the plasmon-LO mode propagated along [100] direction for various concentrations of nonequllibrium carriers $n = p$ and electron temperatures $T_e = 4.2$ K (solid lines) and $T_e = 77$ K (dashed lines) for 6 nm wide HgTe/CdTe QW (a) and 5 nm wide HgTe/Cd$_{0.7}$Hg$_{0.3}$Te QW (b).

## 3.2 Anisotropy

Due to the anisotropy of the hole energy-momentum law, the dispersion of the plasmon-LO mode depends on the direction of propagation. To illustrate this statement, Fig. 5a shows the dispersion laws for the plasmon-LO modes propagating along the [100] and [03-1] directions for a 6 nm wide HgTe/CdTe QW. For other directions of propagation, dispersion lines are located between the lines shown in Fig. 5. It can be seen from the figure that for a fixed value of the wave vector, the frequency of the plasmon-LO mode propagating along the [03-1] direction is slightly higher than the frequency of the mode propagating along other directions (the energy difference is of the order of 1 meV at a fixed $q$).

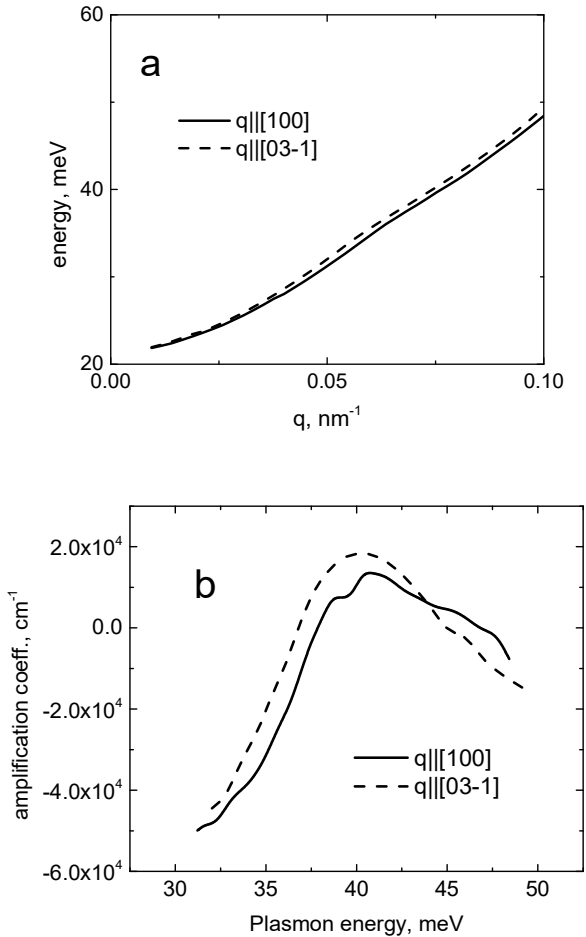

Fig. 5. Dependence of the energy of the plasmon-LO mode on the wave vector (a) and gain spectra for two directions of propagation in a 6 nm wide HgTe/CdTe QW. The concentration of nonequilibrium carriers is $10^{11}$ cm$^{-2}$, $T_e$ = 4.2 K.

Therefore, the dispersion curve of the plasmon-LO mode propagating along the [03-1] direction intersects the dependence $E_{geff}(q)$ at lower values of $q$ as compared to modes propagating in other directions. For this reason, for the plasmon-LO mode propagating along the [03-1] direction, the threshold carrier concentration at which amplification is possible is minimal. A similar situation is realized in the 5 nm wide HgTe/Cd$_{0.7}$Hg$_{0.3}$Te QW.

Fig.5b shows the gain spectra of the plasmon-LO modes propagating along the [100] and [03-1] directions in a 6 nm wide HgTe/CdTe QW. It is clearly seen from the figure that the amplification spectral range for the mode propagating along [03-1] is shifted to lower energies and the magnitude of the gain for this direction in the most part of the amplification range exceeds that for direction [100]. Therefore, the most favorable direction for the generation and amplification of plasmon-LO modes in QWs grown on the (013) plane is the [01-3] direction, at least for carrier concentrations slightly higher than the critical one.

## 4. Conclusion

To conclude, we have shown that the interaction of two-dimensional plasmons with optical phonons leads to the formation of low-frequency and high-frequency modes, similar to the three-dimensional case. The frequency dispersion of the polarizability of the electron and hole gases leads to a nearly linear dependence $\omega(q) \sim q$ for the high-frequency plasmon–LO phonon mode, which can be amplified in the case of an inverted band population. For 6 nm wide HgTe/CdTe QW and 5 nm wide HgTe/Cd$_{0.7}$Hg$_{0.3}$Te QW the gain spectra of the plasmon-LO mode have been calculated for various concentrations of nonequilibrium electron-hole gas, the gain range covering the *Reststrahlen* band of GaAs, where existing quantum cascade lasers do not work. The most favorable direction for amplification of plasmon-LO modes in HgTe/CdHgTe quantum wells grown on the (013) plane is shown to be [03-1] one. Giant values of plasmon-LO mode gain (up to $10^5$ cm$^{-1}$) opens a possibility of creating of compact and effective sources of far-infrared radiation.

## Acknowledgements

This work was supported by the Russian Science Foundation (RSF-ANR Grant # 20-42-09039) and by the French Agence Nationale pour la Recherche (Colector project). The authors are grateful to V.V. Rumyantsev for helpful discussions.

## Appendix. Formula for polarizability of electron gas in a quantum well

Let us consider the effect of an electric field of an electromagnetic wave on an electron gas. Let the electric field present in the sample create an electric potential

$$\varphi(\mathbf{r},t) = \varphi_0 \exp(i\mathbf{q}\mathbf{r} - i\omega t) \quad (1.1)$$

This potential leads to the appearance of an additive to the potential energy of the form

$$U(\mathbf{r},t) = e\varphi_0 \exp(i\mathbf{q}\mathbf{r} - i\omega t) \quad (1.2)$$

Matrix element of a transition between electron state with wavevector k in *s*-th subband and that with wavevector k' in *s'*-th subband equals:

$$U_{\mathbf{k},s;\mathbf{k}',s'} = \exp(-i\omega t)\int \exp(i\mathbf{q}\mathbf{r})\psi^*_{\mathbf{k},s}(\mathbf{r})\psi_{\mathbf{k}',s'}(\mathbf{r})d^3r.$$

Wave functions can be represented as:

$$\psi_{\mathbf{k},s}(\mathbf{r}) = \frac{\exp(i\mathbf{k}\mathbf{r})}{\sqrt{S}}\psi_{\mathbf{k},s}(z), \quad \psi_{\mathbf{k}',s'} = \frac{\exp(i\mathbf{k}'\mathbf{r})}{\sqrt{S}}\psi_{\mathbf{k}',s}(z) \quad (1.3)$$

where $S$ is the quantum well square. In this case, the matrix element is nonzero only when $\mathbf{k}' = \mathbf{k} - \mathbf{q}$ and

$$U_{\mathbf{k},s;\mathbf{k}-\mathbf{q},s'} = e\varphi_0 \int \psi^+_{\mathbf{k},s}(z)\psi_{\mathbf{k}-\mathbf{q},s'}(z)dz = e\varphi_0 \psi^+_{\mathbf{k},s}\psi_{\mathbf{k}-\mathbf{q},s'} \quad (1.4)$$

The presence of a potential leads to the appearance of off-diagonal components of the density matrix, which in the "tau" approximation can be written in the form:

$$\rho_{s,s'}(\mathbf{k},\mathbf{k}-\mathbf{q}) = \frac{e\varphi_0(f_{s'}(\mathbf{k}-\mathbf{q}) - f_s(\mathbf{k}))\psi^+_{\mathbf{k},s}\psi_{\mathbf{k}-\mathbf{q},s'}e^{-i\omega t}}{(i\hbar v_{\mathbf{k},s;\mathbf{k}-\mathbf{q},s'} + \hbar\omega - \varepsilon_s(\mathbf{k}) + \varepsilon_{s'}(\mathbf{k}-\mathbf{q}))} \quad (1.5)$$

where $f_s(\mathbf{k}) = \rho_{s,s}(\mathbf{k},\mathbf{k})$ is the electron distribution function in *s*-th subband, $v_{\mathbf{k},s;\mathbf{k}-\mathbf{q},s'}$ is the phase relaxation frequency. Eq. (1.5) one can receive from equation for off diagonal element of the density matrix (see, for example, [27]). The induced charge density is:

$$\sigma = \frac{e}{S}\sum_{\mathbf{k},s,s'}\rho_{s,s'}(\mathbf{k},\mathbf{k}-\mathbf{q})\psi_{\mathbf{k},s}\psi^+_{\mathbf{k}-\mathbf{q},s'}\exp(i\mathbf{q}\mathbf{r}) =$$
$$= \frac{e^2\varphi_0}{S}\sum_{k}\frac{(f_{s'}(\mathbf{k}-\mathbf{q}) - f_s(\mathbf{k}))|\psi^+_{\mathbf{k},s}\psi_{\mathbf{k}-\mathbf{q},s'}|^2}{(i\hbar v_{\mathbf{k},s;\mathbf{k}-\mathbf{q},s'} + \hbar\omega - \varepsilon_s(\mathbf{k}) + \varepsilon_{s'}(\mathbf{k}-\mathbf{q}))} \quad (1.6)$$

The polarization satisfies the following relation $-\text{div}\mathbf{P} = \sigma$, from which in the considered case we obtain $\mathbf{P} = i\mathbf{q}\sigma/q^2$. On the other hand, the potential is related to the magnitude of the electric field: $\mathbf{E} = -\nabla\varphi = -i\mathbf{q}\varphi$. Using these relations, we find the relationship between the polarization and the electric field:

$$\mathbf{P} = -\frac{e^2\mathbf{E}}{q^2 S}\sum_{\mathbf{k},s,s'}\frac{(f_{s'}(\mathbf{k}-\mathbf{q}) - f_s(\mathbf{k}))|\psi^+_{\mathbf{k},s}\psi_{\mathbf{k}-\mathbf{q},s'}|^2}{(\hbar v_{\mathbf{k},s;\mathbf{k}-\mathbf{q},s'} + \hbar\omega - \varepsilon_s(\mathbf{k}) + \varepsilon_{s'}(\mathbf{k}-\mathbf{q}))} = \chi\mathbf{E} \quad (1.7)$$

From (1.7) we find expression for electron polarizability

$$\chi = \frac{e^2}{q^2 S} \sum_{\mathbf{k},s,s'} \frac{f_s(\mathbf{k}) - f_{s'}(\mathbf{k-q})}{\left(i\hbar v_{\mathbf{k},s;\mathbf{k-q},s'} + \hbar\omega - \varepsilon_s(\mathbf{k}) + \varepsilon_{s'}(\mathbf{k-q})\right)} |\psi^+_{\mathbf{k},s}\psi_{\mathbf{k-q},s'}|^2 =$$

$$= \frac{e^2}{q^2 S} \sum_{\mathbf{k},s,s'} \frac{f_s(\mathbf{k}) - f_{s'}(\mathbf{k+q})}{\left(\varepsilon_{s'}(\mathbf{k+q}) - \varepsilon_s(\mathbf{k}) - \hbar\omega - i\hbar v_{\mathbf{k+q},s';\mathbf{k},s}\right)} |\psi^+_{\mathbf{k+q},s'}\psi_{\mathbf{k},s}|^2$$

(1.8)

that corresponds to Eq. (1) of the present paper. For $v_{\mathbf{k+q},s';\mathbf{k},s} \to 0$ the same result can be obtained from Schrodinger equation using first order perturbation theory.